\documentclass{article}

\usepackage{PRIMEarxiv}

\usepackage{scalerel,stackengine}

\usepackage[utf8]{inputenc} 
\usepackage[T1]{fontenc}    
\usepackage{hyperref}       
\usepackage{url}            
\usepackage{booktabs}       
\usepackage{amsfonts}       
\usepackage{nicefrac}       
\usepackage{microtype}      
\usepackage{lipsum}
\usepackage{fancyhdr}       
\usepackage{graphicx}       
\usepackage{amssymb}
\graphicspath{{media/}}     

\title{Differential Privacy in Federated Learning: Mitigating Inference Attacks with Randomized Response
}

\author{
  Ozer Ozturk,  \\
  Department of Computer Engineering \\
  Marmara University \\
  Istanbul, Turkiye\\
  \texttt{ozer.ozturk@marun.edu.tr} \\
  \And
  Busra Buyuktanir \\
  Department of Computer Engineering \\
  Marmara University \\
  Istanbul, Turkiye\\
  \texttt{busra.buyuktanir@marmara.edu.tr} \\
   \And
  Gozde Karatas Baydogmus \\
  Department of Computer Science \\
  Loyola University of Chicago \\
  Chicago, USA\\
  \texttt{gkaratasbaydogmus@luc.edu} \\
  \And
  Kazim Yildiz \\
  Department of Computer Engineering \\
  Marmara University \\
  Istanbul, Turkiye\\
  \texttt{kazim.yildiz@marmara.edu.tr} \\ \\
}

\begin{document}
\maketitle

\begin{abstract}
Machine learning models used for distributed architectures consisting of servers and clients require large amounts of data to achieve high accuracy. Data obtained from clients are collected on a central server for model training. However, storing data on a central server raises concerns about security and privacy. To address this issue, a federated learning architecture has been proposed. In federated learning, each client trains a local model using its own data. The trained models are periodically transmitted to the central server. The server then combines the received models using federated aggregation algorithms to obtain a global model. This global model is distributed back to the clients, and the process continues in a cyclical manner. Although preventing data from leaving the clients enhances security, certain concerns still remain. Attackers can perform inference attacks on the obtained models to approximate the training dataset, potentially causing data leakage. In this study, differential privacy was applied to address the aforementioned security vulnerability, and a performance analysis was conducted. The Data-Unaware Classification Based on Association (duCBA) algorithm was used as the federated aggregation method. Differential privacy was implemented on the data using the Randomized Response technique, and the trade-off between security and performance was examined under different epsilon values. As the epsilon value decreased, the model accuracy declined, and class prediction imbalances were observed. This indicates that higher levels of privacy do not always lead to practical outcomes and that the balance between security and performance must be carefully considered.
\end{abstract}


\section{Introduction}
The rapid increase in the number of internet-connected devices has led to a significant increase in data production \cite{murshed2021machine}. Sources such as the Internet of Things (IoT), mobile applications, and social media platforms generate large amounts of data at any given moment. Interpreting this large amount of data is becoming increasingly difficult, necessitating the use of machine learning-based approaches \cite{merenda2020edge}. Machine learning models require large amounts of data to achieve sufficient accuracy and performance. Processing big data not only requires high processing power and storage capacity, but also brings various challenges such as data integrity, access speed, security, and privacy.

In distributed systems consisting of clients and servers, traditional machine learning approaches involve collecting data from clients and transferring it to central servers, where model training is performed on this data \cite{liu2020systematic}. In these structures, the requirement to transfer, store, and process large amounts of data on central servers brings significant costs. However, big data structures pose significant risks in terms of privacy protection and data security [3]. As a result of privacy and security breaches, organizations experience not only financial losses but also a loss of reputation and trust. Individuals' increasing sensitivity to personal privacy and the ethical issues that arise as a result make this approach difficult to sustain and increase implementation costs.

\begin{figure}[htp]
    \centering
        \scalebox{0.50}{\includegraphics{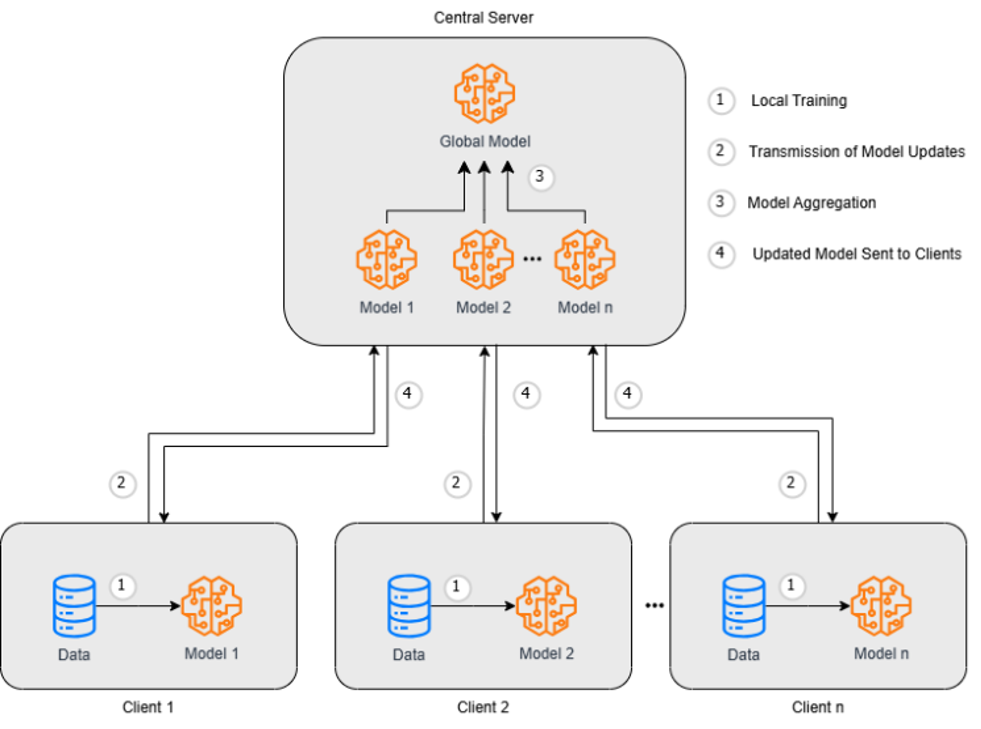}}
    \caption{How the federated learning architecture works}
    \label{fig:figure1}
\end{figure}

As a solution to the aforementioned problems, the federated learning (FL) architecture was developed in 2016 \cite{McMahan2017}. Federated learning is a machine learning approach that eliminates the need for centralized data collection and enables model training using data stored locally on clients in distributed systems. This method is based on the principle of periodically sending models trained locally on clients to a central server, where a global model is created using an appropriate federated aggregation algorithm. The global model is then redistributed to the clients, and the process continues iteratively. Thus, the need for data collection and storage on a central server is eliminated, preserving privacy and enhancing data security. A schematic representation of this architecture is shown in Figure ~\ref{fig:figure1}. 

Although FL architecture has eliminated data sharing and created a more secure system, security concerns remain. There is no complete control over the data, and the control mechanism is weakened due to the distributed architecture. This situation has given rise to a new type of attack called inference attacks \cite{Bagdasaryan2020}. In inference attacks, attackers test the models they have obtained with specific input data and can make inferences about the training data based on the model's outputs \cite{Fredrikson2015}. Furthermore, recent literature also investigates machine unlearning techniques as a way to address privacy and compliance challenges in federated learning systems \cite{ Buyuktanir2025b}. This allows them to predict, with high accuracy, what data the model was trained on and even the contents of some sensitive data samples. This situation weakens federated learning's claim to protect data privacy and raises the need for additional security measures.

This study aims to evaluate the performance of a federated learning architecture by integrating security mechanisms. To this end, the Data-Unaware Classification Based on Association (duCBA) \cite{Buyuktanir2023} method was chosen as the aggregation algorithm, and a prototype was developed to demonstrate the feasibility of the proposed structure. The dataset used in the study was structured as client data; the Randomized Response (RR) method was applied to ensure privacy on the client side. Thus, while privacy was preserved at the client level, the effects on model accuracy were analyzed, and the privacy-performance trade-off was examined in detail.

The rest of this study is structured as follows: The second section reviews the relevant literature. The third section details the methods and data set used in the study. The fourth section shares the findings, while the fifth section discusses the results, addressing the study's contributions, limitations, and suggestions for future research. 

\section{Related Work}
The high communication costs, data privacy issues, and sustainability challenges associated with centralized learning have necessitated the development of new approaches in machine learning. In this context, the federated learning architecture proposed by Google enables local model training on distributed clients without the need for centralized data collection, thereby ensuring both privacy protection and reduced communication load \cite{McMahan2017, Buyuktanir2025a}. In the federated learning process, the parameters of the models trained on the clients are transmitted to the central server at specific intervals; an updated global model is created by merging these models on the server and redistributed to the clients. Thus, the process continues iteratively.

In federated learning systems, model performance largely depends on the effectiveness of federated aggregation algorithms \cite{Onlu2025}. These algorithms merge the updated model parameters received from clients according to specific strategies to produce a new global model. The main algorithms highlighted in the literature are as follows:

\begin{itemize}
    \item \textbf{FedAvg (Federated Averaging):} It is the most commonly used aggregation method. The model weights trained on the clients are averaged to create a global model, weighted according to the amount of data on each client. However, this approach can cause performance fluctuations in cases of heterogeneous (non-IID) data distribution \cite{ McMahan2017, Reddi2020}.
    \item \textbf{FedPer (Federated Learning with Personalization Layers):} It allows only the lower (base) layers of the model to be shared, while the upper (decision) layers are kept on the client. This enables personalization, but the generalization ability of the global model may be limited \cite{Sannara2021}.
    \item \textbf{FedMA (Federated Matched Averaging):} The layer-based neuron matching approach groups similar neurons together. This provides a more structured and meaningful combination; however, this method is notable for its high computational cost \cite{Sannara2021}.
    \item \textbf{FedDist (Federated Distance):} By analyzing the differences between neurons in clients using Euclidean distance, it adds neurons showing significant deviations to the global model. This preserves client-specific information while also increasing the generalization capacity of the global model. With dynamic structure updates, the model architecture can evolve during communication rounds \cite{Sannara2021}.
    \item \textbf{FedSVM (Federated Support Vector Machine):} It is a federated learning algorithm that enables training SVM models on local data residing on distributed clients and merging the weights of these models on a central server. This method allows for updating a global classification model without the need to share data\cite{Nair2022}. 
    \item \textbf{duCBA:} The duCBA aggeration algorithm used in this study is an adaptation of the traditional Classification Based on Associations (CBA) \cite{Chen2006, Tosun2025} algorithm for federated learning environments. duCBA enables clients to generate class-labeled association rules based on their local data, which are then fused on a central server to create a global classification model \cite{Buyuktanir2023}.
\end{itemize}

Although federated learning systems offer a more secure structure in terms of data privacy compared to classical machine learning methods, it cannot be said that security is completely guaranteed \cite{Ciplak2025, Buyuktanir2025c}. These systems are still vulnerable to various types of attacks. In the literature, attacks targeting federated learning are primarily examined under two main categories: model poisoning attacks \cite{Erdol2024} and model inference attacks \cite{Bai2024}.

Model poisoning attacks are attacks where the attacker aims to reduce the accuracy of the global model or steer it in their own interests by injecting models that have been deliberately manipulated into the federated learning process. In such attacks, malicious clients deliberately generate harmful updates and incorporate these updates into the central aggregation process \cite{Erdol2024}.

In contrast, model inference attacks aim to obtain information about the data on which the model was trained \cite{Bai2024}. In this context, two main types of attacks stand out:
\begin{enumerate}
    \item \textit{Membership Inference Attack:} In this type of attack, the attacker attempts to infer whether a specific data sample is included in the training set of the target machine learning model. This is typically achieved by analyzing the uncertainty level in the model's output (e.g., class probabilities or entropy). Overfitting of the model to the training data increases its susceptibility to this attack \cite{Hu2022}.
    \item \textit{Model Inversion Attack:} This attack aims to reconstruct typical input data belonging to a class based on the outputs a model gives to specific classes. It is usually implemented by iteratively optimizing the input data to maximize class probabilities. This method is mostly supported by gradient-based optimization techniques or latent space backprojection methods \cite{Dibbo2023}.
\end{enumerate}

One of the primary defense approaches developed against these attacks is the principle of differential privacy \cite{Xiong2020}. Differential privacy is a privacy definition that mathematically guarantees that third parties cannot reliably determine whether an individual is included in a dataset. This principle can be implemented through the Laplace \cite{Phan2017} and Gauss \cite{Liu2018} mechanisms. Both mechanisms are designed for numerical data and are difficult to apply directly to categorical data. The Randomized Response method \cite{Ye2025}, a suitable alternative for categorical data, was first proposed by Warner in 1965 and has been used to protect the privacy of individuals in survey studies. With approaches developed today, the Randomized Response method can be widely applied to ensure privacy on categorical data within the framework of differential privacy.

\section{Methodology}
This section elaborates on the methods and methodology employed in the study.

\subsection{Technologies Used and Development Environment}
This study was developed using the Python programming language, and the Spyder integrated development environment (IDE) included in the Anaconda distribution was preferred. The application was implemented using the pyarc, numpy, pandas, scikit-learn, and pyfim libraries. These libraries were effectively used in data preprocessing, inference of association rules, and classification processes.

\subsection{Dataset}
In this study, the hypertension subset of the “Diabetes, Hypertension and Stroke Prediction” dataset, compiled by Prosper Chuks and made available on the Kaggle data sharing platform, was used for hypertension prediction \cite{Chuks2025}. The dataset is based on the 2015 Behavioral Risk Factor Surveillance System (BRFSS) survey conducted by the Centers for Disease Control and Prevention (CDC) in the United States and has been made available to researchers with preprocessing steps completed.

The subset used contains approximately 26,000 samples and 14 attributes. The class distribution for the target variable, hypertension, was observed to be 45.3\% (n $=$ 11,809) positive and 54.7\% (n $=$ 14,274) negative. The Imbalance Ratio (IR $\approx$ 1.21) calculated based on this distribution is below the commonly accepted threshold value of 1.5 in the literature, so the dataset is considered balanced \cite{Kraiem2021}. Furthermore, there are no missing observations in the dataset, and it is structured in a way that can be directly applied to machine learning models. Table ~\ref{tab:table1} presents summary information about the dataset used in the study.

\begin{table}[htp]
\centering
\caption{Information about dataset's features}
\label{tab:table1}
\begin{tabular}{lll}
\hline
\multicolumn{1}{c}{\textbf{Column Name}} & \multicolumn{1}{c}{\textbf{Data Type}} & \multicolumn{1}{c}{\textbf{Explanation}}                        \\ \hline
age                                      & Numerical                              & Age                                                             \\
sex                                      & Categorical                            & Gender                                                          \\
cp                                       & Categorical                            & Type of chest pain                                              \\
trestbps                                 & Numerical                              & Resting blood pressure                                          \\
chol                                     & Numerical                              & Serum cholesterol                                               \\
fbs                                      & Categorical                            & A condition where fasting blood sugar is greater than 120 mg/dl \\
restecg                                  & Categorical                            & Resting ECG results                                             \\
thalach                                  & Numerical                              & Maximum heart rate                                              \\
exang                                    & Categorical                            & Exercise-induced chest pain                                     \\
oldpeak                                  & Numerical                              & Exercise-induced ST segment depression                          \\
slope                                    & Categorical                            & The slope of the ST segment during exercise                     \\
ca                                       & Categorical                            & Number of large coronary arteries visualized by fluoroscopy     \\
thal                                     & Categorical                            & Myocardial perfusion status determined by thallium stress test  \\
target                                   & Categorical                            & The presence or absence of hypertension                         \\ \hline
\end{tabular}
\end{table}

\subsection{Data Preprocessing}
To ensure the model training was conducted properly, various data preprocessing steps were applied within the scope of this study. First, observations containing missing values were removed from the dataset, thereby preventing deviations due to missing data. Since the duCBA algorithm can only work with categorical data, numerical variables were appropriately converted into categorical form. In this context, for the thalach (maximum heart rate) variable, which was not suitable for direct use, a new derived variable named thalach$\_$ratio was created by utilizing the correlation between this variable and age. This variable was divided into specific intervals and converted into a categorical format.

The chi-square test of independence was applied to determine the relationship between the independent variables in the dataset and the target variable \cite{Yakar2024}. As a result of this test, the variables age and gender, which did not show a statistically significant relationship with the target variable, were removed from the dataset. Thanks to these preprocessing steps, the data set was made suitable for the requirements of the duCBA algorithm, and structural integrity was ensured prior to model training.

\subsection{Federated Learning Architecture and duCBA Merging Algorithm}
In this study, a methodology based on a federated learning architecture has been designed for edge devices such as mobile devices, IoT systems, and client-server-based collaborative structures, and within this scope, the duCBA federated aggregation algorithm has been applied. For the duCBA algorithm to function, clients use the CBA algorithm with their local data to generate models based on classification rules. The CBA algorithm operates within the scope of supervised data mining, producing labeled association rules based on specific support and confidence threshold values. Once trained, the local models are transmitted by the clients to the central server only as structures consisting of labeled rules; the data itself is not sent to the server.

These models collected on the server are integrated through a merging module specific to the duCBA algorithm \cite{Buyuktanir2023}. This module compares the rules received from clients in terms of content and labels, updates the support and confidence values of identical rules, and prefers the rule with the higher support value when matching rules have different labels. In case of equality, priority is given to the rule that arrived first. All rules are ranked according to their confidence value; in case of equality, a secondary ranking is applied based on support value and then order. The resulting rule list represents the final model created on the server. In this process, special formulas have been developed to update the support and confidence values according to their weighted average.

Within the scope of this study, a simulation environment was created to evaluate the duCBA algorithm from the perspectives of privacy and security. The dataset used was divided into two parts: 80\% training and 20\% testing. The training data was divided into equal parts, assuming it came from different clients, and the simulation of the endpoints was performed. For example, in a scenario with two clients, the training set was randomly divided into two approximately equal parts. Each part was trained separately using the CBA algorithm, and only the generated models (i.e., labeled rules) were sent to the server. The final model was obtained by merging these rules using the duCBA algorithm.

During model training, the support and confidence thresholds were set at 0.02 and 0.5, respectively. The final model was evaluated on the test data and a performance analysis was performed.

\subsection{Differential Privacy}
Within the duCBA architecture, each unit on the client side uses its local data to generate models based on classification rules through the CBA algorithm. However, during this process, the anonymity of client data is at risk from privacy threats such as model inference attacks. As a countermeasure against this threat, a differential privacy mechanism has been integrated into the client side.

In this study, the Randomized Response method, which is particularly effective on categorical data for differential privacy applications, has been chosen. This method is applied to client data prior to model training to enable the concealment of sensitive information. It aims to prevent the identification of individual participants by randomly distorting the accuracy of each data sample based on the principle of randomness.

The implementation of the differential privacy mechanism can negatively affect the classification performance of the model by compromising data integrity to a certain extent. In this context, there is a trade-off between the level of privacy and model performance. This trade-off is controlled through the epsilon ($\epsilon$) value, which is the differential privacy parameter. As $\epsilon$ value decreases, the privacy level in the system increases; however, this increase can lead to a decrease in accuracy rates by increasing the amount of randomness integrated into the model.

Within the scope of the study, a simulation environment consisting of three clients was created. A privacy mechanism was applied to each client's data, and local model training was performed. These local models were then merged on the server side to obtain a global model. This process was repeated with different $\epsilon$ values to analyze the impact of the privacy-performance trade-off on the system. The findings revealed that this trade-off must be carefully managed.

\section{Experimental Results}
This section presents the experimental findings obtained. The classification performance of the developed model was evaluated using fundamental metrics such as accuracy, precision, recall, and F1 score, both with and without privacy enhancement.  

\begin{figure}[htp]
    \centering
        \scalebox{0.50}{\includegraphics{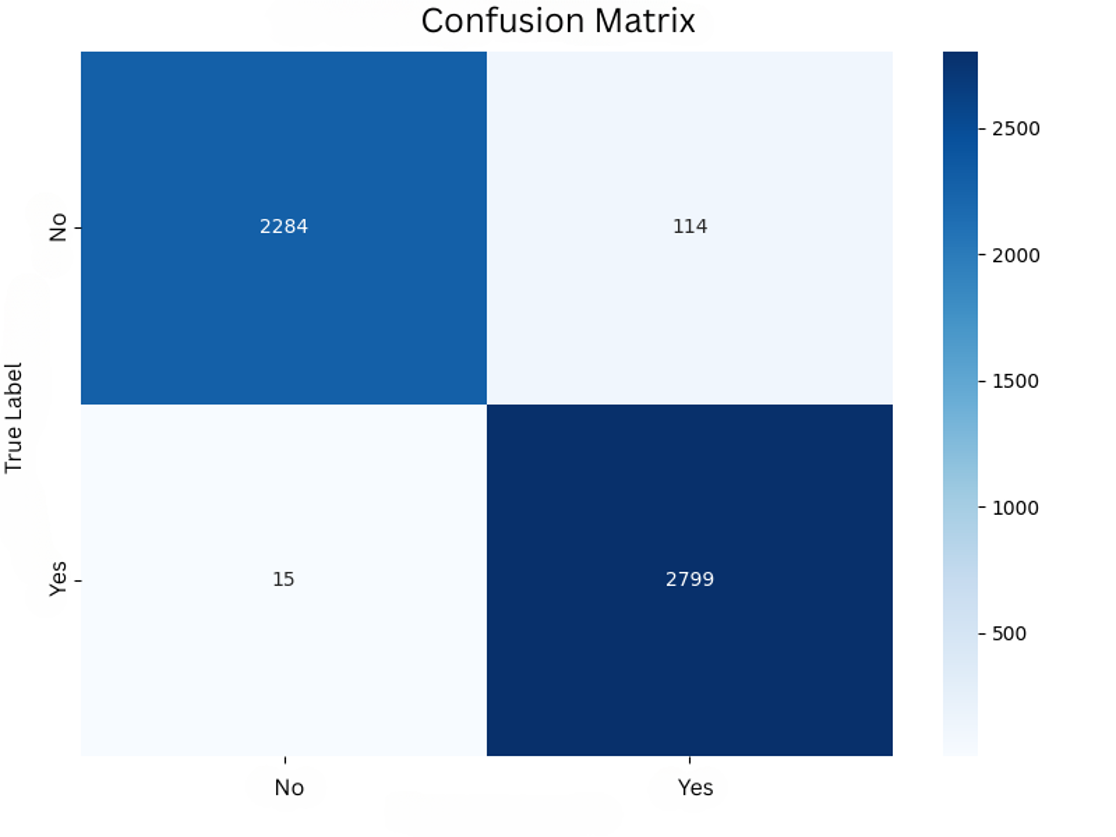}}
    \caption{Confusion matrix of the final merged-duCBA model}
    \label{fig:figure2}
\end{figure}

\begin{figure}[htp]
    \centering
        \scalebox{0.50}{\includegraphics{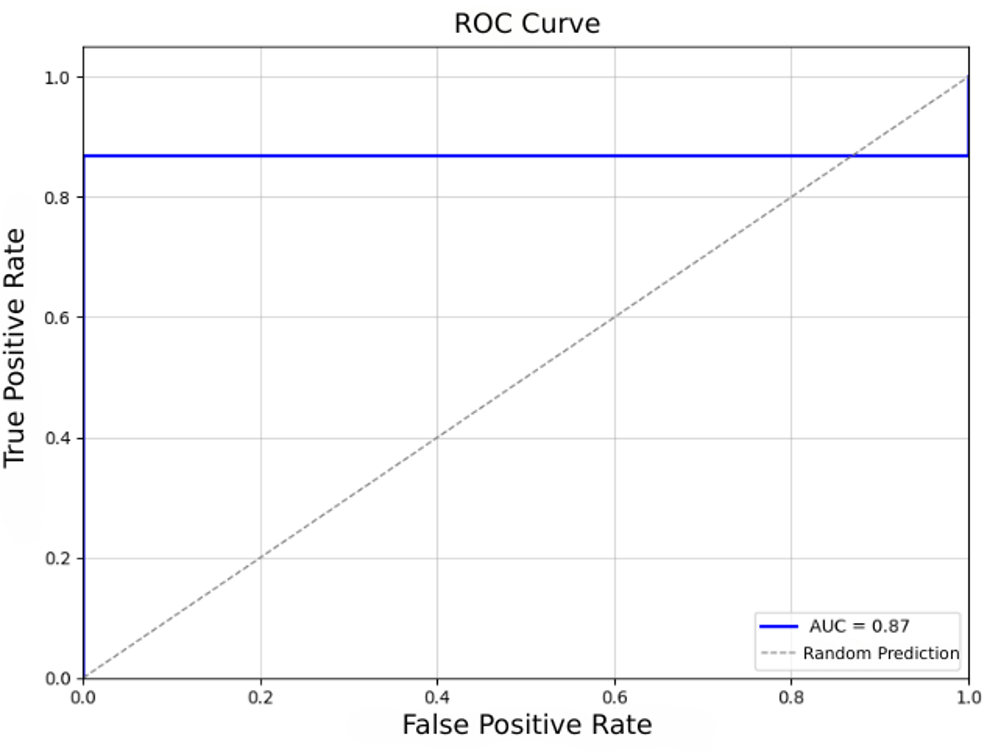}}
    \caption{ROC curve of the duCBA merged final model}
    \label{fig:figure3}
\end{figure}

The model was evaluated within a structure simulating the FL architecture, and the work was carried out locally on a single computer. In this context, a federated learning scenario was applied, acting as if there were a decentralized structure. First, the dataset was split into two parts: 80\% for training and 20\% for testing. The training data was randomly divided into three equal parts, and each part was configured to represent an independent client. On each client, local model training was first performed using the CBA algorithm. Then, these locally trained models on the clients were merged using the duCBA aggregation algorithm to obtain a centralized global model. The results of the performance evaluation of the merged model obtained without privacy enhancement are presented in Table ~\ref{tab:table2}, the confusion matrix in Figure ~\ref{fig:figure2}, and the ROC curve in Figure ~\ref{fig:figure3}.

\begin{table}[htp]
\centering
\caption{Performance metrics for the proposed model without RR}
\label{tab:table2}
\begin{tabular}{lllll}
\hline
Class           & Precision & Recall & F1-Score & Accuracy \\ \hline
No Hypertension & 99.00     & 95.00  & 97.00    & 97.00    \\
Hypertension    & 96.00     & 99.00  & 98.00    & 97.00    \\
Macro Average   & 98.00     & 97.00  & 98.00    & 97.00    \\ \hline
\end{tabular}
\end{table}

The accuracy rate of the merged final model was calculated as 97\%. This high accuracy rate indicates that the model generally demonstrated successful classification performance. When examining the confusion matrix in Figure ~\ref{fig:figure2}, it can be seen that 2284 examples were correctly classified as true negatives, while 2799 examples were correctly classified as true positives; on the other hand, only 114 examples were classified as false positives and 15 examples were classified as false negatives. Figure ~\ref{fig:figure3} shows the general form of the ROC curve and an AUC value of 87\%, indicating that the model exhibits high discriminative power between classes and demonstrates reliable performance in terms of generalization.
     
Based on these results, it can be said that the model is particularly successful in distinguishing between hypertension classes. Furthermore, the fact that the F1 scores calculated on a class basis are both high and close to each other indicates that the model was able to learn both classes in a balanced manner and perform classification without bias towards any class. This also demonstrates that the model's power of discrimination between classes is strong and reliable. 

Subsequently, model performance was evaluated by integrating a privacy mechanism into the same structure. First, $\epsilon$ value was set to 1, and the performance results of the obtained model were examined. Local model training was performed using the CBA algorithm, and differential privacy was applied to the client data during this process. These locally trained models were merged on the central server using the duCBA merging algorithm to obtain a global model. The performance evaluation results of this merged model with the privacy mechanism applied are shown in Table ~\ref{tab:table3}{}.

\begin{table}[htp]
\centering
\caption{RR model's performance metrics}
\label{tab:table3}
\begin{tabular}{lllll}
\hline
Class           & Precision & Recall & F1-Score & Accuracy \\ \hline
No Hypertension & 94.00     & 65.00  & 77.00    & 83.00    \\
Hypertension    & 78.00     & 97.00  & 86.00    & 83.00    \\
Macro Average   & 86.00     & 81.00  & 82.00    & 83.00    \\ \hline
\end{tabular}
\end{table}

Table ~\ref{tab:table2} shows that the model without privacy applied demonstrated high F1 scores and balanced performance for both classes. However, in Table ~\ref{tab:table3}, when differential privacy (RR) was added, there was a significant decrease in sensitivity and F1 scores, particularly in the “No Hypertension” class. which indicates that the privacy mechanism has an asymmetric effect on the classes.

\begin{figure}[htp]
    \centering
        \scalebox{0.50}{\includegraphics{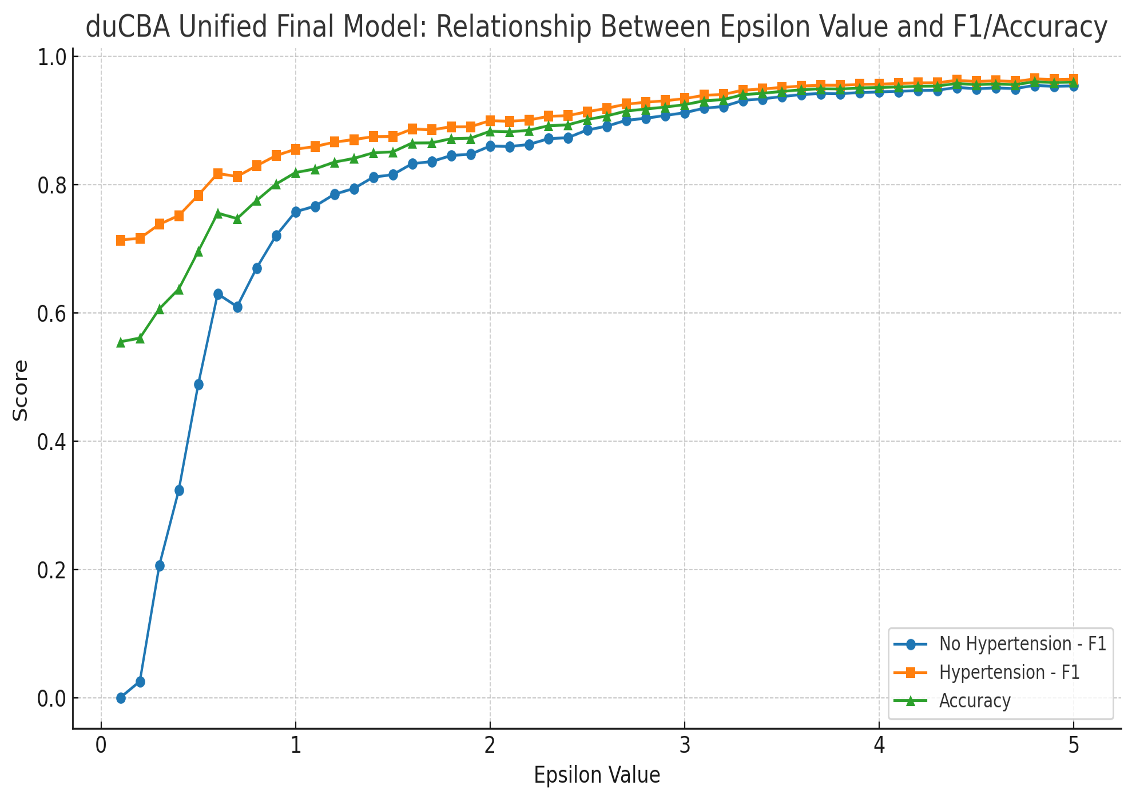}}
    \caption{Model accuracy and class-based F1 scores according to the Epsilon value}
    \label{fig:figure4}
\end{figure}

The changes in model performance were analyzed by repeating the process for different values of $\epsilon$ between 0 and 5. Figure ~\ref{fig:figure4} graphically shows the changes observed in model accuracy and class-based F1 scores as $\epsilon$ value increases. As $\epsilon$ value decreased, i.e., as the differential privacy level increased, a decrease was observed in both the overall accuracy rate and the F1 scores for both classes.

Specifically, the F1 score for the “No Hypertension” class was more affected and produced lower results compared to the “Hypertension Present” class at low $\epsilon$ values. This situation indicates that the randomization process applied within the scope of differential privacy may have an asymmetric effect on class patterns.

In the literature, it is accepted that $\epsilon$ <= 1 provides strong confidentiality according to the NIST SP 800-226 guide. On the other hand, it is stated that confidentiality guarantees weaken as the $\epsilon$ value increases. However, the general recommendation is to choose the lowest possible $\epsilon$ value to the extent that performance loss is tolerable.

Figure ~\ref{fig:figure4} shows the changes observed in model accuracy and class-based F1 scores as the $\epsilon$ value increases. As $\epsilon$ value decreases, etc., as the differential privacy level increases, a decrease is observed in both the overall accuracy rate and the F1 scores for both classes. This reveals that the noise operation applied to preserve privacy significantly reduces the model's classification performance, especially at low $\epsilon$ values. The F1 score for the "No Hypertension" class was more affected at low $\epsilon$ values and produced lower scores compared to the "Hypertension Present" class. This indicates that the effect of the noise process on class patterns may be asymmetric. In the literature, $\epsilon$ <= 1 is considered to provide strong privacy according to the NIST SP 800$-$226 guide. It is accepted that privacy guarantees weaken as the $\epsilon$ value increases. However, as a general recommendation, the lowest possible $\epsilon$ value should be used as long as the performance loss is tolerable.

\section{Discussion and Conclusion}
Within the scope of this study, the duCBA aggregation algorithm, developed based on a federated learning architecture, has been evaluated from the perspectives of privacy and security. The duCBA algorithm is an adaptation of the traditional CBA method for federated environments, enabling the integration of classification rules trained locally on clients into a centralized structure. However, merely preventing centralized data sharing in federated learning structures does not provide complete assurance against privacy violations. Therefore, differential privacy mechanisms have been utilized, particularly to strengthen individual data privacy against inference attacks.

In the study, the Randomized Response method was applied within the scope of differential privacy, and model updates were anonymized by adding controlled noise to the data generated on the client side. In the experimental evaluations conducted, an inverse relationship was observed between the applied noise level and model accuracy; this revealed that $\epsilon$ parameter plays a critical role in system performance. As $\epsilon$ value decreases, the level of privacy increases, but this increase leads to a significant decrease in the model's classification performance. The findings obtained indicate that the privacy-performance trade-off must be carefully analyzed in the design of federated systems.

Future work will focus on developing security measures against model poisoning attacks, another threat to federated learning systems, and increasing the resilience of the duCBA algorithm against such attacks.

\bibliographystyle{unsrt}  

\begin{thebibliography}{10}

\bibitem{murshed2021machine}
MG~Sarwar Murshed, Christopher Murphy, Daqing Hou, Nazar Khan, Ganesh Ananthanarayanan, and Faraz Hussain.
\newblock Machine learning at the network edge: A survey.
\newblock {\em ACM Computing Surveys (CSUR)}, 54(8):1--37, 2021.

\bibitem{merenda2020edge}
Massimo Merenda, Carlo Porcaro, and Demetrio Iero.
\newblock Edge machine learning for ai-enabled iot devices: A review.
\newblock {\em Sensors}, 20(9):2533, 2020.

\bibitem{liu2020systematic}
Yi~Liu, Li~Zhang, Ning Ge, and Guanghao Li.
\newblock A systematic literature review on federated learning: From a model quality perspective.
\newblock {\em arXiv preprint arXiv:2012.01973}, 2020.

\bibitem{McMahan2017}
B.~McMahan, E.~Moore, D.~Ramage, S.~Hampson, and B.~A. y~Arcas.
\newblock Communication-efficient learning of deep networks from decentralized data.
\newblock In {\em Proc. Int. Conf. Artificial Intelligence and Statistics (AISTATS)}, pages 1273--1282, Apr 2017.

\bibitem{Bagdasaryan2020}
E.~Bagdasaryan, A.~Veit, Y.~Hua, D.~Estrin, and V.~Shmatikov.
\newblock How to backdoor federated learning.
\newblock In {\em Proc. Int. Conf. Artificial Intelligence and Statistics (AISTATS)}, pages 2938--2948, Jun 2020.

\bibitem{Fredrikson2015}
M.~Fredrikson, S.~Jha, and T.~Ristenpart.
\newblock Model inversion attacks that exploit confidence information and basic countermeasures.
\newblock In {\em Proc. 22nd ACM SIGSAC Conf. Computer and Communications Security (CCS)}, pages 1322--1333, Oct 2015.

\bibitem{Buyuktanir2025b}
B.~Buyuktanir, K.~Yildiz, and G.~K. Baydogmus.
\newblock A systematic mapping study on machine unlearning in federated learning.
\newblock In {\em Proc. 7th Int. Congress on Human-Computer Interaction, Optimization and Robotic Applications (ICHORA)}, pages 1--10. IEEE, May 2025.

\bibitem{Buyuktanir2023}
B.~Büyüktanir, K.~Yildiz, E.~Ülkü, and T.~Bütüktanir.
\newblock du-cba: Data-agnostic and incremental classification-based association rules extraction architecture.
\newblock {\em Journal of the Faculty of Engineering and Architecture of Gazi University}, 38(3), 2023.

\bibitem{Buyuktanir2025a}
B.~Buyuktanir, {\c{S}}.~Altinkaya, G.~Karatas Baydogmus, and K.~Yildiz.
\newblock Federated learning in intrusion detection: advancements, applications, and future directions.
\newblock {\em Cluster Computing}, 28(7):1--25, 2025.

\bibitem{Onlu2025}
A.~{\"O}. {\"O}nl{\"u}, B.~Akca, B.~Buyuktanir, K.~Yildiz, and G.~K. Baydogmus.
\newblock An investigation and performance evaluation of aggregation algorithms in federated learning architecture.
\newblock In {\em Proc. Innovations and Applications in Smart Systems Conf. (ASYU)}, Istanbul, Turkey, 2025.

\bibitem{Reddi2020}
S.~Reddi, Z.~Charles, M.~Zaheer, Z.~Garrett, K.~Rush, J.~Konečný, and H.~B. McMahan.
\newblock Adaptive federated optimization.
\newblock arXiv preprint arXiv:2003.00295, Mar 2020.

\bibitem{Sannara2021}
E.~K. Sannara, F.~Portet, P.~Lalanda, and V.~E. G.~A. German.
\newblock A federated learning aggregation algorithm for pervasive computing: Evaluation and comparison.
\newblock In {\em Proc. IEEE Int. Conf. Pervasive Computing and Communications (PerCom)}, pages 1--10, Mar 2021.

\bibitem{Nair2022}
D.~G. Nair, C.~V.~A. Narayana, K.~J. Reddy, and J.~J. Nair.
\newblock Exploring svm for federated machine learning applications.
\newblock In {\em Advances in Distributed Computing and Machine Learning, Proc. ICADCML}, pages 295--305, Singapore, 2022. Springer.

\bibitem{Chen2006}
G.~Chen, H.~Liu, L.~Yu, Q.~Wei, and X.~Zhang.
\newblock A new approach to classification based on association rule mining.
\newblock {\em Decision Support Systems}, 42(2):674--689, 2006.

\bibitem{Tosun2025}
B.~E. Tosun, E.~Yorulmaz, F.~E. Ergul, B.~Buyuktanir, G.~K. Baydogmus, and K.~Yildiz.
\newblock Comparative analysis of relational classification algorithms on benchmark datasets.
\newblock In {\em Proc. 10th Int. Conf. Computer Science and Engineering (UBMK)}, Istanbul, Turkey, 2025.

\bibitem{Ciplak2025}
Z.~{\c{C}}{\i}plak, K.~Y{\i}ld{\i}z, and {\c{S}}.~Alt{\i}nkaya.
\newblock Fedetect: a federated learning-based malware detection and classification using deep neural network algorithms.
\newblock {\em Arabian Journal for Science and Engineering}, pages 1--28, 2025.

\bibitem{Buyuktanir2025c}
B.~Buyuktanir, Z.~Ciplak, A.~E. Cil, O.~Yakar, M.~B. Adoum, and K.~Yildiz.
\newblock Ddos\_fl: Federated learning architecture approach against ddos attack.
\newblock {\em Pamukkale University Journal of Engineering Sciences}, 31(6), 2025.

\bibitem{Erdol2024}
E.~S. Erdöl, H.~Erdöl, B.~Üstübioğlu, F.~Z. Solak, and G.~Ulutaş.
\newblock Impactful neuron-based secure federated learning.
\newblock In {\em Proc. 32nd Signal Processing and Communications Applications Conf. (SIU)}, pages 1--4, May 2024.

\bibitem{Bai2024}
L.~Bai, H.~Hu, Q.~Ye, H.~Li, L.~Wang, and J.~Xu.
\newblock Membership inference attacks and defenses in federated learning: A survey.
\newblock {\em ACM Computing Surveys}, 57(4):1--35, 2024.

\bibitem{Hu2022}
H.~Hu, Z.~Salcic, L.~Sun, G.~Dobbie, P.~S. Yu, and X.~Zhang.
\newblock Membership inference attacks on machine learning: A survey.
\newblock {\em ACM Computing Surveys}, 54(11s):1--37, 2022.

\bibitem{Dibbo2023}
S.~V. Dibbo.
\newblock Sok: Model inversion attack landscape: Taxonomy, challenges, and future roadmap.
\newblock In {\em Proc. 36th IEEE Computer Security Foundations Symposium (CSF)}, pages 439--456, Jul 2023.

\bibitem{Xiong2020}
X.~Xiong, S.~Liu, D.~Li, Z.~Cai, and X.~Niu.
\newblock A comprehensive survey on local differential privacy.
\newblock {\em Security and Communication Networks}, 2020:8829523, 2020.

\bibitem{Phan2017}
N.~Phan, X.~Wu, H.~Hu, and D.~Dou.
\newblock Adaptive laplace mechanism: Differential privacy preservation in deep learning.
\newblock In {\em Proc. IEEE Int. Conf. Data Mining (ICDM)}, pages 385--394, Nov 2017.

\bibitem{Liu2018}
F.~Liu.
\newblock Generalized gaussian mechanism for differential privacy.
\newblock {\em IEEE Transactions on Knowledge and Data Engineering}, 31(4):747--756, 2018.

\bibitem{Ye2025}
Q.~Ye, L.~Yu, K.~Huang, X.~Xiao, W.~Liu, and H.~Hu.
\newblock From randomized response to randomized index: Answering subset counting queries with local differential privacy.
\newblock In {\em Proc. IEEE Symposium on Security and Privacy (SP)}, pages 3877--3891, May 2025.

\bibitem{Chuks2025}
P.~Chuks.
\newblock Diabetes, hypertension and stroke prediction [dataset].
\newblock \url{https://www.kaggle.com/datasets/prosperchuks/health-dataset}, 2025.
\newblock Accessed: Jul. 18, 2025.

\bibitem{Kraiem2021}
M.~S. Kraiem, F.~S{\'a}nchez-Hern{\'a}ndez, and M.~N. Moreno-Garc{\'i}a.
\newblock Selecting the suitable resampling strategy for imbalanced data classification regarding dataset properties: An approach based on association models.
\newblock {\em Applied Sciences}, 11(18):8546, 2021.

\bibitem{Yakar2024}
O.~Yakar, B.~Buyuktanir, A.~E. Cil, and A.~B.~A. Girgin.
\newblock Performance comparison of different classification algorithms and feature selection methods in turkish hate speech problem analysis.
\newblock {\em European Journal of Science and Technology}, (53):97--111, 2024.

\end{thebibliography}

\end{document}